\documentclass{aa}  

\usepackage{graphicx}
\usepackage{txfonts}
\begin{document}

   \title{Star formation in a diffuse high-altitude cloud?}

   \author{J. Kerp
          \and
        D. Lenz
\and
T. R\"ohser
          }

   \institute{Argelander-Institut f\"ur Astronomie
              Auf dem H\"ugel 71, D-53121 Bonn, Germany\\
              \email{jkerp@astro.uni-bonn.de}
             }

   \date{Received September 15, 1996; accepted March 16, 1997}

 
  \abstract
   {A recent discovery of two stellar
clusters associated with the diffuse high-latitude cloud HRK 81.4--77.8 has important implications for star formation in the Galactic halo.
}
   {We derive a plausible distance estimate to HRK 81.4-77.8 primarily from its gaseous properties.
   }
   {We spatially correlate state-of-the-art HI, far-infrared and soft X-ray data to analyze the diffuse gas in the cloud. The absorption of the soft X-ray emission from the Galactic halo by HRK 81.4-77.8 is used to constrain the distance to the cloud.
   }
   {HRK 81.4-77.8 is most likely located at an altitude of about 400 pc within the disk-halo interface of the Milky Way Galaxy. 
The HI data discloses a disbalance in density and pressure between the warm and cold gaseous phases. Apparently, the cold gas is compressed by the warm medium. This disbalance might trigger the formation of molecular gas high above the Galactic plane on pc to sub-pc scales.
   }
   {}


   \maketitle
%

\section{Introduction}
\citet{Camargo2015} report on the discovery of two embedded stellar clusters proposed to be physically associated with the high Galactic latitude cloud HRK\,81.4-77.8 \citep[or][G\,81.4-77.8]{Heiles1988}.
Using WISE data \citep{Wright2010} \citet{Camargo2015} deduce from their color-magnitude diagrams (CMDs) an age of about 2\,Myr for both stellar clusters, Camargo\,438 and Camargo\,439. Based on the absolute stellar luminosity, a distance from the Sun of about 5\,kpc is calculated. Because of their location close to the southern Galactic pole they are positioned nearly at the solar circle, but at an altitude of about 4\,kpc below the Galactic plane.
Moreover, \citet{Camargo2015} identified two cluster member stars with known proper motions towards Camargo\,439. This information is used to estimate the orbital parameters of the extra-planar clusters.

Our aim is to explore the physical conditions and the interstellar gas properties of HRK\,81.4-77.8. So far only two high Galactic latitude clouds that form stars
at altitudes of a few tens of pc above the Galactic plane, namely MBM\,12 and MBM\,20 \citep{MBM1985}, have been identified  \citep[see][]{McGehee2008}. Both star-forming MBM clouds are classified as dark interstellar clouds. Dark interstellar clouds show up with an optical extinction in excess of $A_{\rm V} = 5$\,mag \citep{McGehee2008}.

We use state-of-the-art multifrequency data for our investigation. The HI 21 cm line data of the Parkes Galactic All--Sky Survey (GASS) \citep{KalberlaHaud2015, Kalberla2010, McClure2009} and the Planck far-infrared (FIR) data \citep{PlanckXI2014} are analyzed to quantify the amount of molecular gas in HRK\,81.4-77.8. The interstellar extinction is evaluated by the updated \citep{ SchlaflyFinkbeiner2011} Schlegel, Finkbeiner \& Davis (SFD) data set \citep{Schlegel1998}. ROSAT all--sky survey data \citep{SnowdenSchmitt1990, Snowden1997} are analyzed to evaluate the soft X-ray shadow of HRK\,81.4-77.8 allowing to estimate its distance from the Sun.
Section\,2 comprises the details of our analyses, while Sect.\,3 contains our conclusions.

\section{Gaseous properties of HRK\,81.4-77.8}
\subsection{HRK\,81.4-77.8: A diffuse infrared cirrus cloud}
Using the improved SFD data set \citep{Schlegel1998, SchlaflyFinkbeiner2011} on interstellar extinction  HRK\,81.4-77.8 reveals a maximum of only $A_{\rm V} = 0.144$\,mag. According to this value it has to be classified as a diffuse cloud \citep{McGehee2008}. 
Diffuse interstellar clouds are not expected to  efficiently shield molecules against dissociation by high-energy photons, even within their densest portions. Thus, significant fractions of molecular gas are not expected to be present under these environmental conditions \citep{KennicuttEvans2012}.
Hence, the physical association of the stellar clusters Camargo 438 and 439 with HRK\,81.4-77.8 is a striking proposal.
\subsection{HRK\,81.4-77.8: Distance estimate derived from interstellar reddening}
Very recently \citet{Green2015} published a 3-D database of interstellar dust reddening based on optical and near-infrared surveys. Inspecting this data cube for the extinction towards HRK\,81.4--77.8 discloses a faint signal. The best fit extinction shows a continuous increase in the reddening up to a distance of 400\,pc. Beyond that distance the reddening remains constant out to 15\,kpc. When   HRK\,81.4--77.8 is located at a distance of 5\,kpc  we would expect a second increase in the reddening at this particular distance, which is not observed. Accordingly, the \citet{Green2015} database is not consistent with a location of HRK\,81.4--77.8 at 5\,kpc.

   \begin{figure*}
     \centerline{
        \includegraphics[height=5cm, angle=0]{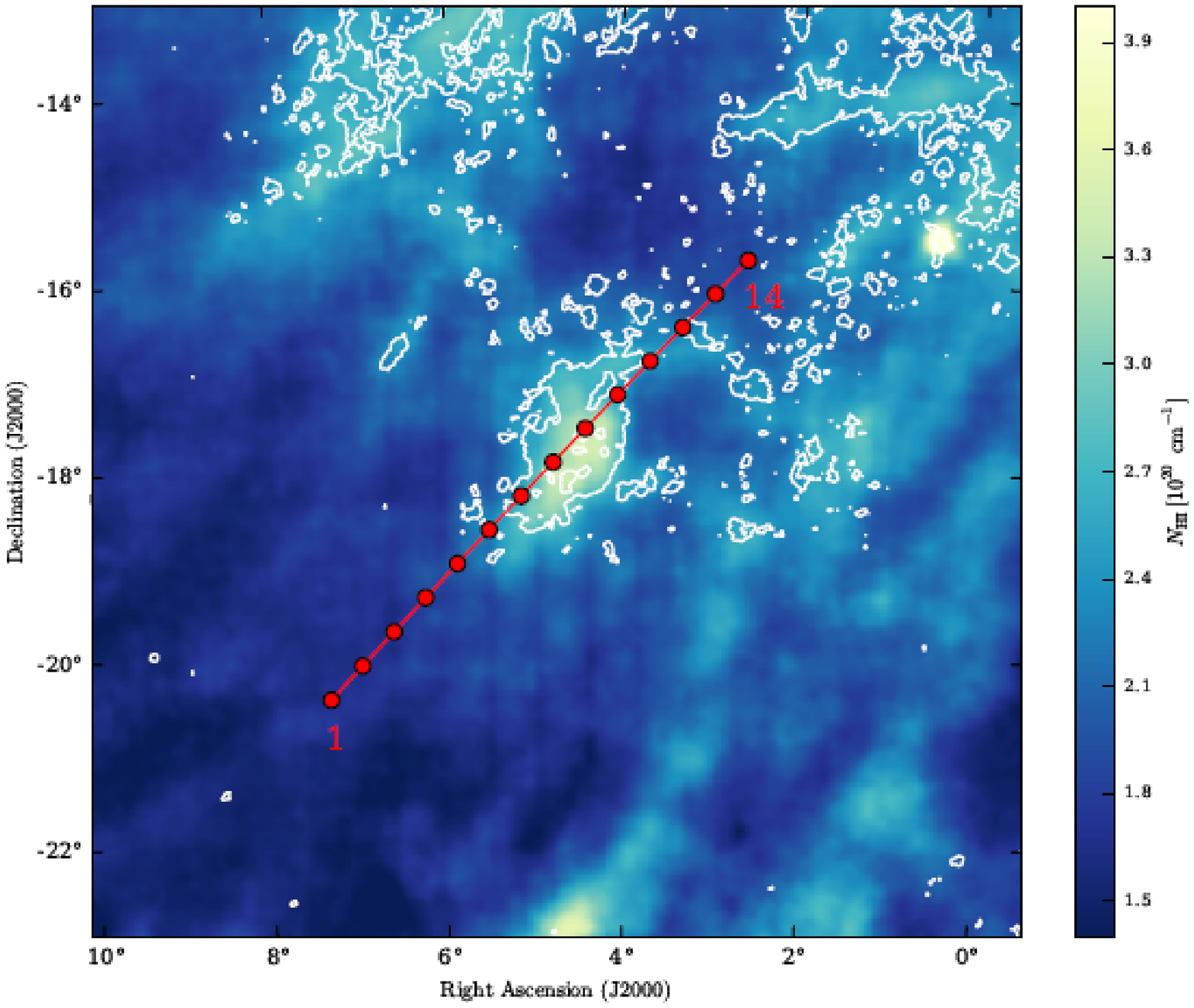}
        \includegraphics[height=5cm , angle=0]{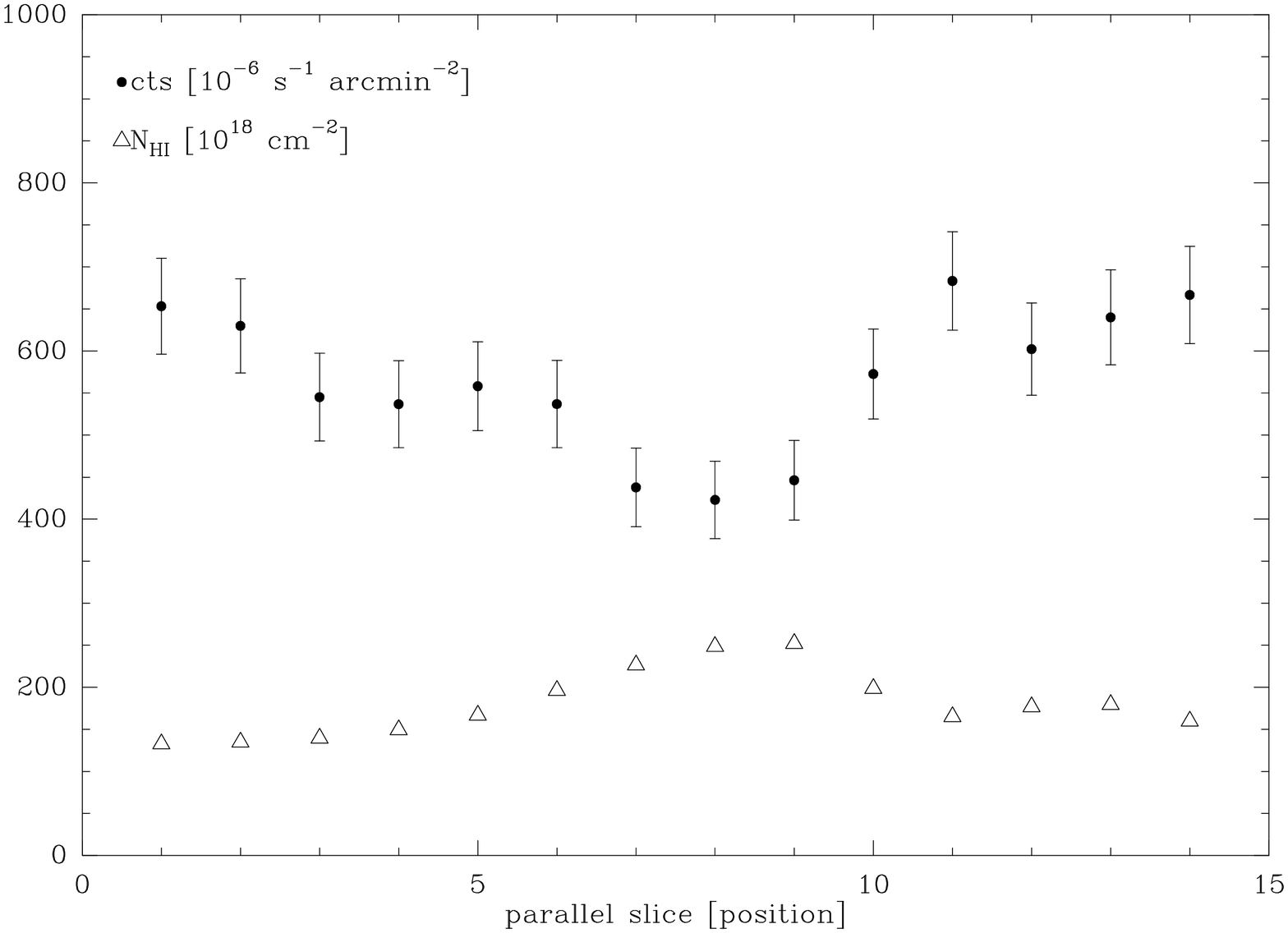}
                }
      \caption{{\bf Left:} Integrated ($-37.3\,\leq\,v_{\rm LSR} {[\rm km\,s^{-1}]}\,\leq\,52.9$) GASS HI column density map of HRK\,81.4--77.8. Superposed as contours is the Planck 857\,GHz intensity distribution starting at 0.76\,${\rm MJy\cdot sr^{-1}}$ in steps of 0.45\,${\rm MJy\cdot sr^{-1}}$. 
{\bf Right:} Slice across the longest extent of HRK\,81.4--77.8 (see left panel, dots along the solid line). The open triangles mark the HI column density distribution $N_{\rm HI}$ in $10^{18}\,{\rm cm^{-2}}$, the filled circles represent the ROSAT R12 count rate in $10^{-6}\,{\rm cts\,s^{-1}\,arcmin^{-2}}$.}
    \label{Fig:mom0}
   \end{figure*}

\subsection{Estimating the distance to HRK\,81.4-77.8: ROSAT All-Sky Survey data}
Soft X-rays with photon energies below the carbon edge of 0.284\,keV are very sensitive to photoelectric absorption by the interstellar gas \citep{Wilms2000}.

The effective photoelectric absorption cross section at a column density of approximately $N_{\rm HI} = 3.0\cdot10^{20}\,{\rm cm^{-2}}$ is $\overline{\sigma} \simeq 0.8 \cdot 10^{-20}\,{\rm cm^2}$ for a thermal plasma of $T = 10^{6.2}$\,K \citep[see][Fig.\,4]{Snowden1994}. Thus, even diffuse interstellar gas at high galactic latitude modulates the soft X-ray background intensity distribution significantly.

Prior to the ROSAT mission \citep{SnowdenSchmitt1990, Snowden1997} it was thought that the observed soft X-ray background emission originates entirely from a thermal plasma \citep{Snowden1990} filling the local cavity \citep{FrischYork1983, Paresce1984, Lallement2014}.
In consequence, soft X-ray brightness variations are solely due to different path lengths through the coronal plasma, sampling essentially the shape of the local bubble \citep{Snowden1990, McCammonSanders1990}.

The discovery of a soft X-ray shadow of the high galactic altitude Draco cloud \citep{Snowden1991} led to a revision of the local hot bubble model. Moreover, \citet{Pietz1998} demonstrated that the soft X-ray background intensity distribution can be modeled quantitatively across the full sky adopting a simplified soft X-ray radiation transfer: 
\begin{equation}
I = I_{\rm local} + I_{\rm distant}\cdot e^{-\overline{\sigma}\cdot N_{\rm HI}}
\label{Eq:radtrans}
.\end{equation}
In addition to the local soft X-ray thermal emission, $I_{\rm local}$ a much more luminous and slightly hotter thermal plasma component $I_{\rm distant}$ fills a flattened Milky Way halo (scale heights 4.4\,kpc, scale length 15\,kpc; \citet{Pietz1998}). This distant soft X-ray emission originates beyond the bulk of the neutral hydrogen \citep[see][Fig.\,11]{KalberlaKerp2009}.
Consequently, \citet{Snowden2000} released a catalog of 378 soft X-ray shadows associated with individual HI clouds.

Their Table\,1 contains as shadow  \#100 a four degree extended soft X-ray absorption feature positionally coincident with HRK\,81.4-77.8.
In Fig.\,\ref{Fig:mom0} we show the $N_{\rm HI}$ map of HRK\,81.4-77.8 (left-hand panel) and a slice across the whole cloud (right-hand panel). The ROSAT R12 count rates and the HI column density disclose a positional negative correlation. Accordingly, HRK\,81.4-77.8 is physically coincident with the soft X-ray shadow  \#100 of \citet{Snowden2000}.
The maximum HI column density of HRK\,81.4-77.8 along that slice is $N_{\rm HI} \simeq 2.5\cdot 10^{20}\,{\rm cm^{-2}}$. Calculating the product of $N_{\rm HI}$ and the effective soft X-ray absorption cross section yields $\overline{\sigma}\cdot N_{\rm HI} = 1.8$ for the ROSAT R12 band.
An attenuation of about 83\% of all soft X-ray emission originating beyond HRK\,81.4-77.8 is expected (see Eq.\ref{Eq:radtrans}). In summary, towards the high column density portion of the cloud we observe dominantly $I_{\rm local}$, while the soft X-ray shadow detection confirms by itself a localization of the cloud in front of the bulk of the Milky Way halo plasma.
\citet{Pietz1998} determined that the scale height of the Milky Way X-ray halo is about 4.4\,kpc. Adopting a distance of 500\,pc about 50\% of the halo plasma is located beyond HRK\,81.4-77.8. Thus, the {\it ROSAT\/} soft X-ray shadow implies a nearby location of HRK\,81.4-77.8 because of its significant depth. However, an upper distance limit of 500\,pc is compatible with the cloud's absorption feature within the uncertainties of the X-ray data.

Towards the deepest portions of the soft X-ray shadow of HRK\,81.4-77.8 we observe dominantly $I_{\rm local}$. This foreground emission is a measure for the minimum separation between the cloud and the Sun.
\citet{Snowden2000} determined an $I_{\rm local}$ count rate of only $(3.7\pm 0.4)\cdot 10^{-4}\,{\rm cts\,s^{-1}\,arcmin^{-2}}$ in the ROSAT R12 energy band.
This is quantitatively equal to the $I_{\rm local}$ count rate observed towards MBM\,12 \citep{SnowdenMBM12} with $(3.7\pm 0.4)\cdot 10^{-4}\,{\rm cts\,s^{-1}\,arcmin^{-2}}$. 
However, MBM12 is located at a distance of $(360\pm 30$)\,pc \citep{Andersson2002}. 

Because of systematic uncertainties in determining the absolute quantity of $I_{\rm local}$ \citep{Galeazzi2014, Puspitarini2014} we explore the error--range of this distance estimate.
For this aim we inspect the local count rate towards
\begin{itemize}
\item the Draco cloud (\citet{Snowden2000} shadow \#116, $I_{\rm local} = (4.8\pm 0.2)\cdot 10^{-4}\,{\rm cts\,s^{-1}\,arcmin^{-2}}$ at a distance between 463\,pc and 618\,pc \citet{Gladders1998}) and
\item IVC\,135+54 (\citet{Snowden2000} shadow \#182, $I_{\rm local} = (5.8\pm 0.2)\cdot 10^{-4}\,{\rm cts\,s^{-1}\,arcmin^{-2}}$ at $(395\pm 95)$\,pc \citet{Benjamin1996}).
\end{itemize}
Both $I_{\rm local}$ count rates are in excess of that observed towards HRK\,81.4-77.8. Implying that HRK\,81.4-77.8 is most likely located in between MBM\,12 and both IVCs within a distance bracket of 300\,pc to 500\,pc.
While the lower distance limit is affected by systematic uncertainties in evaluating $I_{\rm local}$ \citep{Galeazzi2014}, the depth of the strong soft X-ray shadow sets a robust upper distance limit, which is  of greater importance here for our discussion.
Adopting this distance bracket we have to consider HRK\,81.4-77.8 as a low-extinction, diffuse interstellar cloud located within the Milky Way's disk-halo interface. 
Thus, our result is in conflict with the 5\,kpc distance estimate of \citet{Camargo2015}. Consequently, the stars of both stellar clusters might be significantly fainter and of much lower mass than initially assumed.

\subsection{Estimating the distance to HRK\,81.4-77.8: Proper motion information towards Camargo\,439}
Associated with the stellar cluster Camargo\,439 are two proper motion measurements; the absolute length of the proper motion vectors are both about 6.2\,marcsec/year \citep[see][Tab.\,2]{Camargo2015}. This proper motion corresponds to
\begin{eqnarray} 
6.2\cdot 10^{-3}\,{\rm arcsec\,year^{-1}} & \equiv & 29.3\,{\rm km\,s^{-1}\,kpc^{-1}} 
\end{eqnarray} 
yielding $v_\bot \simeq 147\,{\rm km\,s^{-1}}$ at a distance of 5\,kpc, as \citet{Camargo2015} assume. These very high tangential velocities are unexpected for young stellar clusters \citep{Foster2015}, suggesting that Camargo\,439 needs to be extremely young to be detected as an entity because the two stellar proper motion vectors are oriented almost perpendicular to each other. \citet{Camargo2015} estimated an age of $2\pm 1$\,Myr for Camargo\,439 which appears to be incompatible with the high proper motion. Moving for a period of 1 Myr with $6\,{\rm marcsec\,year^{-1}}$ would displace the stars by more than a degree from their birthplace. These numbers are inconsistent with the apparent overdensity of the stellar members of Camargo\,439 \citep[see][Fig.\,2 left]{Camargo2015}.

Adopting the hypothesis that Camargo\,439 is physically associated with HRK\,81.4-77.8 and consequently located at an upper distance limit of 400\,pc yields $v_\bot \simeq 11.7\,{\rm km\,s^{-1}}$. Even this value is in some conflict with actual studies of the stellar velocity dispersion of a young stellar cluster \citep{Foster2015}, but of the same order of magnitude as the turbulence broadened HI line width of the warm gas component of HRK\,81.4-77.8 (Fig.\,\ref{Fig:HIspectrum}). 

   \begin{figure}
     \centerline{
        \includegraphics[width=5cm, angle=270]{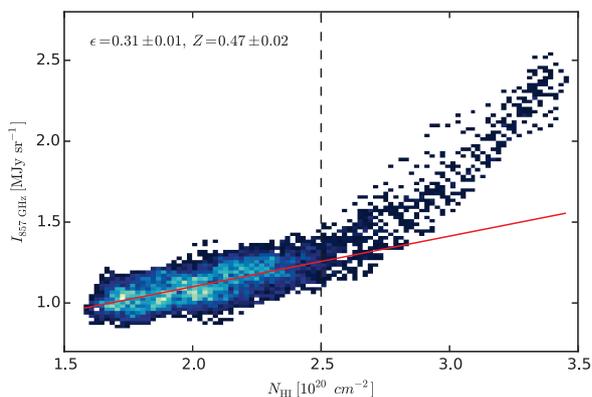}
            }
      \caption{
$N_{\rm HI}$ versus $I_{\rm FIR}$ 857\,GHz correlation plot. The solid line represents a linear approximation (see Eq. \ref{Eq:FIR}) to the data. Above $N_{\rm HI} = 2.5\cdot 10^{20}\,{\rm cm^{-2}}$ FIR excess emission is observed. This emission is attributed to the presence of additional molecular hydrogen regions not traced by $N_{\rm HI}$.}
\label{Fig:scatter857vsHI}
\end{figure}
\subsection{FIR emission of HRK\,81.4-77.8}
\label{HIDustCorrelation}
\citet{Heiles1988} investigated the FIR, HI, and carbon monoxide (CO) properties of 26 isolated, degree-sized interstellar clouds. In the case of HRK\,81.4-77.8 they did not report on CO \citep[see][Tab.\,2]{Heiles1988}. \citet{Magnani2000} performed a southern hemisphere survey of the $^{12}{\rm CO}(1\rightarrow 0)$ transition.  Their survey also does not comprise any information on the CO content of HRK\,81.4-77.8 \citep[see][Fig.\,1]{Magnani2000}.

The maximum HI column density of HRK\,81.4-77.8 at GASS angular resolution of $14.4'$ is $N_{\rm HI} = 3.9\cdot 10^{20}\,{\rm cm^{-2}}$. This neutral hydrogen column density exceeds the molecule formation threshold value of $N_{\rm HI} = 2.0\cdot 10^{20}\,{\rm cm^{-2}}$ \citep{PlanckXXIV2011}. Above that value the FIR emissivity cannot be approximated by $N_{\rm HI}$ alone; additionally, we have to account for molecular hydrogen $N_{\rm H_2}$ because $I_{\rm FIR} \propto N_{\rm H} = N_{\rm HI} + 2\cdot N_{\rm H_2}$. Associated with HRK\,81.4-77.8 we expect the presence of a detectable amount of molecular hydrogen from the FIR.

Correlating the Planck FIR 857\,GHz and the GASS HI data allows us to evaluate the gas--to--dust ratio and the amount of $N_{\rm H_2}$. Molecular gas, not quantitatively traced by neutral hydrogen, can be identified by an excess over the linear FIR-to-$N_{\rm HI}$ correlation. 
Figure\,\ref{Fig:scatter857vsHI} shows the 857\,GHz FIR emission versus the HI column density. For the quantitative evaluation we smooth the Planck data to the lower angular resolution of GASS. The linear portion of the HI/FIR correlation is evaluated by
\begin{equation}
I_{\rm FIR} = \epsilon \cdot N_{\rm HI} + Z
\label{Eq:FIR}
.\end{equation}
For the dust emissivity we find $\epsilon = (0.31\pm 0.01)\cdot 10^{-20}\,{\rm MJy\, sr^{-1}}$ and $Z = (0.47\pm 0.02)\,{\rm MJy\, sr^{-1}}$.
The slope is compatible to the mean Galactic value between $\epsilon = 0.3 - 0.7\cdot 10^{-20}\,{\rm MJy\,sr^{-1}\,cm^2}$ \citep{Boulanger1996, PlanckXXIV2011}. This  too is  a strong argument for a nearby location of HRK\,81.4-77.8, in addition to the ROSAT soft X-ray shadow. HRK\,81.4-77.8 appears to be exposed to the same radiation field as a typical local Galactic interstellar cloud.

\subsection{Molecular gas fraction of HRK\,81.4-77.8}
Although CO has not yet been observed in HRK\,81.4-77.8, CO-dark molecular hydrogen may be present. Figure\,\ref{Fig:scatter857vsHI} shows that above $N_{\rm HI} = 2.0\cdot 10^{20}\,{\rm cm^{-2}}$ the linear relation between dust and gas does not match quantitatively. Excess FIR emission is commonly attributed to the presence of molecular hydrogen $N_{\rm H_2}$ \citep{Reach1998}.
An upper limit for $N_{\rm H_2}$ can be determined by attributing all FIR excess emission to $N_{\rm H_2}$. At a resolution of  14.4\,arcmin  we derive a peak column density for ${\rm H_2}$ of $N_{\rm H_2} = 2.8\cdot 10^{20}\,{\rm cm^{-2}}$ from the FIR/HI correlation shown in Fig.\,\ref{Fig:scatter857vsHI}. We chose for this scatter diagram a circular aperture centered at the HI peak with a radius of 2.7 degrees. Consequently HRK\,81.4-77.8 is dominantly an atomic cloud with a significant fraction of molecular gas. Locally a molecular fraction of up to $f = 60$\% $\left(f = \frac{2\cdot N_{\rm H_2}}{N_{\rm HI} + 2\cdot N_{\rm H_2}}\right)$ is deduced. These regions are positionally coincident with cold neutral medium (CNM) gas.

To estimate the volume-filling factor of the molecular gas in HRK\,81.4-77.8, which are below the angular resolution level of the Parkes telescope,  we evaluate the FIR intensity contrast between the individual FIR peaks and their neighborhood at Planck's full angular resolution of 4.3\,arcmin at 857\,GHz. On these angular scales, corresponding to physical sizes of about 0.5\,pc (at 400\,pc altitude),   a variation in FIR brightness of up to a factor of two can be determined. This suggests, that the volume density is not expected to increase by more than an order of magnitude between linear scales of 0.5\,pc and 1.6\,pc (Parkes resolution). All gaseous properties of HRK\,81.4-77.8 are consistent with normal high galactic latitude clouds. Moreover, the significant fraction of molecular gas can be considered  typical for an infrared cirrus object \citep{GillmonShull2006}.

\subsection{HI properties of HRK\,81.4-77.8}
Its extinction classifies HRK\,81.4-77.8 as a diffuse cloud. It is host to a significant fraction of molecular gas of up to 60\% of the total hydrogen column density. Most likely it is located within the disk--halo interface of the Milky Way Galaxy. A few high galactic latitude clouds at these altitudes have been studied so far, but disclosing very similar physical properties \citep{Moritz1998, Weiss1999, Heithausen2001, Hernandez2013, Roehser2014, Lenz2015}. 
The most significant difference is, however, the association of HRK\,81.4-77.8 with two stellar clusters. If this association is a real one and not an accidental, this would have major implications for the formation of dwarf stars and brown dwarfs.

   \begin{figure}
     \centerline{
        \includegraphics[width=8.5cm, angle=0]{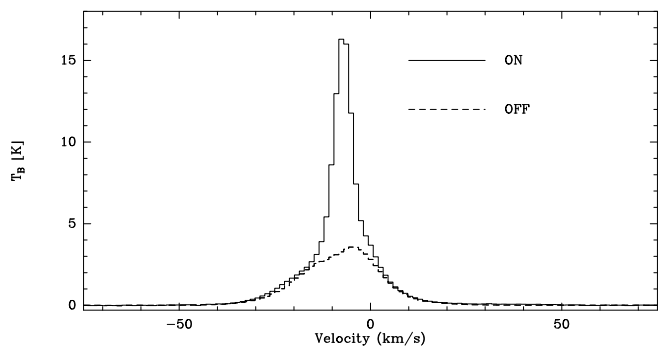}
            }
      \caption{Parkes GASS HI 21-cm line spectrum towards the stellar cluster Camargo\,439 (solid line). Next to the bright narrow HI line a underlying broad HI emission line is visible. Offset from {\bf that position} we extracted an HI spectrum of the unrelated HI emission (dashed line).}
\label{Fig:HIspectrum}
    \end{figure}

\subsubsection{Density perturbation in the warm neutral medium}
\label{Denpert}
Figure\,\ref{Fig:HIspectrum} shows the HI spectrum towards Camargo\,438 averaged across an area of a single GASS beam. It discloses a two-component HI spectrum.
The peak brightness temperature is about $T_{\rm B} = 17.7$\,K and the full width at half maximum (FWHM) is only $\Delta v_{\rm narrow}({\rm FWHM}) = 4.9\,{\rm km\,s^{-1}}$. This narrow line width corresponds to an upper limit for the kinetic temperature of $T_{\rm kin} = 550$\,K. In such a cold cloud the velocity of sound is only $c_{\rm S} = 2.7\,{\rm km\,s^{-1}}$.

The HI line profile displays a second  HI line component that is dimmer but broader. Approximating simultaneously two Gaussian profiles to this HI spectrum yields almost the same Doppler velocity of the CNM $v_{\rm narrow} = -6.9\,\pm 0.1\,{\rm km\,s^{-1}}$ and a warm 
$v_{\rm broad} = -7.2\,\pm 0.1\,{\rm km\,s^{-1}}$ component. The line width of the warm neutral medium (WNM) is $\Delta v_{\rm broad}(FWHM) = 22\,{\rm km\,s^{-1}}$. The column density of the broad component is $N_{\rm HI}(WNM) = 1.8\cdot10^{20}\,{\rm cm^{-2}}$ and about 30\% higher than for the narrow line component ($N_{\rm HI}(CNM) = 1.3\cdot 10^{20}\,{\rm cm^{-2}}$).

Offset from HRK\,81.4-77.8 we extracted an HI spectrum of the unrelated environmental HI gas. It is shown in Fig.\,\ref{Fig:HIspectrum} by the dashed line. This off--cloud HI spectrum fits the second broad HI component towards Camargo\,438 nicely.
This finding implies that the warm HI gas component ON Camargo\,438 is the same as OFF to HRK\,81.4-77.8.

Accordingly, we cannot attribute the whole WNM gas component observed towards Camargo\,438 to HRK\,81.4-77.8, but to the WNM gas layer of the Milky Way Galaxy. The similarity of the radial velocities of HRK\,81.4-77.8 and the WNM argues, however, for a common origin of both. HRK\,81.4-77.8 might be the result of a distortion in the velocity field (ram pressure) of the WNM initiating the molecular gas formation.

\subsubsection{Ram pressure}
Figure\,\ref{Fig:sixpanels} shows the dynamics of the HI gas. Displayed are six brightness temperature maps of subsequent spectral channels. Starting at $v_{\rm LSR} = -12.78\,{\rm km\,s^{-1}}$ it is possible to see that even beyond the FIR emission of HRK\,81.4-77.8 dynamically related HI emission is observed. We note that the displayed Doppler velocity panels sample the narrow (CNM) portion of the HI spectral line of HRK\,81.4--77.8 (Fig.\,\ref{Fig:HIspectrum}).
The whole field of interest is filled with HI gas moving at comparable Doppler velocities and sharing similar properties with respect to spectral line width and brightness temperature to HRK\,81.4-77.8. A remarkable rapid structural variation of the HI brightness temperature from channel to channel is obvious. With decreasing Doppler velocity we can identify a ring like structure in the middle and bottom panels of Fig.\,\ref{Fig:sixpanels}. It encircles the FIR emission of HRK\,81.4-77.8.

Across the extent of the FIR emission of HRK\,81.4-77.8, Fig.\,\ref{Fig:sixpanels} discloses a Doppler velocity gradient of at least $6.4\,{\rm km\,s^{-1}}$. This Doppler velocity gradient will cross the whole cloud within a few $10^5$\,years.

\section{Discussion and conclusions}
In this paper we     determine  an upper distance limit for HRK\,81.4--77.8. This is an important piece of information considering the challenging proposal of \citet{Camargo2015} that the densest portions of this clouds are associated with star formation at an altitude of about 5\,kpc above the Galactic plane.
The low extinction, the FIR emissivity, the implausible tangential velocities, and the soft X-ray shadow consistently argue for a location of HRK\,81.4--77.8 at an altitude of between 300 and 500\,pc. Its Doppler velocity classifies HRK\,81.4--77.8 as a low-velocity cloud with a two-phase medium. HRK\,81.4--77.8 is continuously connected to neighboring HI gas showing up with comparable radial velocities and HI emission properties. 
In all aspects HRK\,81.4--77.8 is a normal high galactic altitude cloud hosting some molecular regions.
These molecular regions are of about 0.5\,pc extent and show up with a total hydrogen mass of $n_{\rm H} = n_{\rm HI} + 2\cdot n_{\rm H_2}$ of $M_{\rm H} \simeq 10\,{\rm M_\odot}$.
The association with the young stellar clusters is unexpected \citep{Camargo2015}. 
Special in addition to HRK\,81.4--77.8 is however the large scale radial velocity gradient, perhaps tracing a volume density/gas pressure situation out of any equilibrium between the WNM and CNM. 
This implies that the cloud might be in a transient state and not in an equilibrium state.

\begin{figure*}
     \centerline{
        \includegraphics[scale=0.6, angle=0]{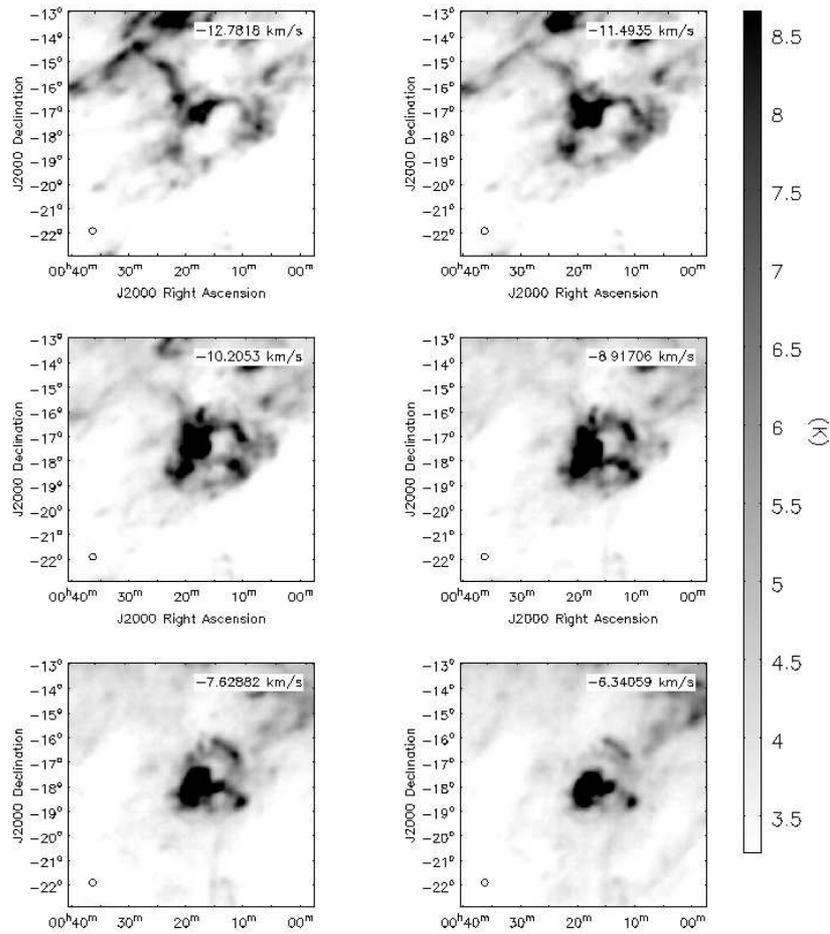}
            }
      \caption{HI 21-cm brightness temperature maps of six subsequent velocity channels. HRK\,81.4-77.8 is located in the very center of the map. Note, that bright HI emission is detected across the whole field of interest, comparable in both, the brightness as well as in line width.  HRK\,81.4-77.8 is hence part of a much larger extended HI structure. Remarkable is the gas motion from top-left to  HRK\,81.4-77.8 in the very center in hand with the formation of a ring-like structure encircling  HRK\,81.4-77.8.}
    \label{Fig:sixpanels}

    \end{figure*}

Figure\,\ref{Fig:sixpanels} is suggestive for a ram--pressure interaction between the gas associated with HRK\,81.4-77.8 and the disk--halo interface gas layers. This ram pressure might provide the necessary additional energy to trigger a WNM to CNM transition similar to the situation described by \citet{Roehser2014} and R\"ohser et al. (2015, submitted). Eventually, this will lead to the formation of molecular hydrogen even towards low column density clouds in the disk-halo interface, as reported by \citet{GillmonShull2006, Planck2011}.

At the altitude of the disk--halo interface (about 400\,pc; \citep{KalberlaKerp2009}, the Galactic radiation field is sufficiently thinned out \citep{Haffner2009}. Hence, the probability of detecting a significant fraction of molecular gas towards low column density diffuse clouds is high. That this is indeed the case, even in low-extinction diffuse interstellar clouds, is observationally confirmed by multiple observations via HI/FIR correlations and CO emission maps \citep{Moritz1998, Weiss1999, Heithausen2001, Roehser2014, Lenz2015}. This holds true also on linear scales below one pc. Here, molecular structures have been observed as a common phenomenon towards the high Galactic latitude sky as reported by \citet{Heithausen2007} and \citet{GillmonShull2006}.

When stars are really formed at high galactic altitudes, they will chemically and dynamically resemble their gaseous nursery. Smooth gradients in the chemical composition of dwarf halo stars and  in their angular momentum distribution should be observed \citep{Fuhrmann1998, Nidever2014},  but first and foremost we need to study the stellar population of Camargo\,438 and 439 in much more detail.

\begin{acknowledgements}
The authors thank the Deutsche Forschungsgemeinschaft (DFG) for supporting  the Effelsberg-Bonn HI Survey project under  grant number KE\,757/7-1 to -3 KE\,757/9-1 and KE\,757/11-1. D.L. and T.R. are members of the Bonn--Cologne Graduate School of Physics and Astronomy (BCGS). T.R. is a member of the International Max--Planck Research School (IMPRS) for Astronomy and Astrophysics at the Universities of Bonn and Cologne.
\end{acknowledgements}

\bibpunct{(}{)}{;}{a}{}{,} 
\bibliographystyle{aa} 
\bibliography{references} 


\end{document}